\documentclass[12pt,a4paper]{article}
\usepackage{graphicx}

\begin{document}



\begin {center}
{\LARGE \textbf{Systematic Study of Two-Pion Production in $NN$ Collisions --- 
from Single-Baryon to Di-Baryon Excitations\\}} 

T. Skorodko, M. Bashkanov, H. Clement, E. Doroshkevich, O. Khakimova,
A. Pricking and G. J. Wagner for the CELSIUS/WASA Collaboration \\

\vspace*{0.5cm}
Physikalisches Institut der Universit\"at T\"ubingen, Auf der Morgenstelle 14,
D-72076 T\"ubingen, Germany

\begin{abstract}
 The two-pion production in nucleon-nucleon collisions has been studied by
 exclusive and kinematically complete experiments from threshold up to $T_p$ = 
 1.36 GeV at CELSIUS-WASA. At near-threshold energies the total and
 differential distributions for the $\pi^+\pi^-$ and $\pi^0\pi^0$ channels are
 dominated by Roper excitation and its decay into $N\sigma$ and $\Delta\pi$ 
 channels. At beam energies $T_p >$ 1.1 GeV the $\Delta\Delta$ excitation
 governs the two-pion production process. In the $\pi^+\pi^+$ channel evidence
 is found for the excitation of a higher-lying I=3/2 resonance, favorably the
 $\Delta(1600)$. The isovector fusion processes leading to the deuteron and to 
 quasi-stable $^2$He, respectively,  
 exhibit no or only a modest ABC-effect, {\it i.e.} low-mass enhancement in the
 $\pi\pi$-invariant mass spectrum, and can be described by conventional
 $t$-channel $\Delta\Delta$ excitation. On the other hand, the isoscalar
 fusion process to the deuteron 
 exhibits a dramatic ABC-effect correlated with a narrow resonance-like energy
 dependence in the total cross section with a width of only 50 MeV and
 situated at a mass 90 MeV below the $\Delta\Delta$ mass. 
\end{abstract}
\end{center}

\section{Introduction}

The $\pi\pi$ production in isoscalar, isovector and isotensor
$\pi\pi$ channels has been systematically studied in nucleon-nucleon ($NN$)
collisions 
at CELSIUS both in double-pionic fusion reactions and in reactions, where the
two participating nucleons do not fuse into a final nuclear bound system.

According to theoretical predictions \cite{luis} the $NN$ initiated two-pion
production should be dominated by excitation and decay of resonances in the
course of the reaction process. In the channels, where two pions can be in the 
scalar-isoscalar state, the main production mechanism at low energies is
expected to be the
Roper excitation. This provides the unique possibility to study the excitation
of the still puzzling Roper resonance and its subsequent decay into the
$N\pi\pi$ system at low incident energies, where no other competing processes
are expected to contribute significantly.

At energies $T_p\ge$ 1.1 GeV the mutual excitation of both nucleons into their
first excited state, the $\Delta$ resonance, is expected to take over. Note
that the latter process can not take place in photo- or pion-induced
single-baryon reactions. Hence the NN collision process is uniquely suited for
the study of the $\Delta\Delta$ excitation, which may have a large impact on
the question regarding the interaction between two excited nucleons.

The two-pion production in double-pionic fusion reactions deals with the
problem of an unexpected low-mass enhancement in the $\pi\pi$ invariant
mass ($M_{\pi\pi}$) spectrum, known as ABC effect. This effect has been an
intriguing puzzle 
all the time since its discovery 50 years ago by Abashian, Booth and
Crowe \cite{abc} and is named after the initials of these authors. 
Follow-up experiments \cite{hom,hall,bar,abd,plo,ban,ban1,wur,col} revealed this
effect to be of isoscalar nature with regard to the $\pi\pi$ system and to
happen solely in 
cases, where the two-pion production process leads to a bound nuclear
system. With the exception of low-statistics bubble-chamber measurements all
previous experiments carried out on this issue have been inclusive
measurements conducted preferentially with single-arm magnetic spectrometers
for the detection of the emitted fused nuclei. Since the ABC effect occured
at beam energies corresponding roughly to the excitation energy of two
$\Delta$s, the ABC effect was interpreted 
\cite{wur,col,ris,barn,anj,gar,alv}  by $t$-channel $\Delta\Delta$ excitation 
in the course of the reaction process leading to both a low-mass and a 
high-mass enhancement in the isoscalar $M_{\pi\pi}$ spectra. In fact, the
missing momentum spectra from inclusive measurements have been in support of
such predictions.

\section{Experiment}

Exclusive measurements of the reactions $pp\rightarrow pp \pi^0\pi^0$,
$pp\rightarrow pp\pi^+\pi^-$, $pp\rightarrow pn\pi^+\pi^0$, $pp\rightarrow
nn\pi^+\pi^+$, $pp\rightarrow d\pi^+\pi^0$ and pd$\rightarrow$pd$\pi^0\pi^0$
have been carried out at beam energies in the range $T_p$ = 0.65 - 1.36 GeV at
the CELSIUS storage ring using the 4$\pi$ WASA detector setup \cite{barg}
including the pellet target system. The pd$\rightarrow$pd$\pi^0\pi^0$ reaction
proceeds as quasifree 
pn$\rightarrow$d$\pi^0\pi^0$ reaction with a spectator proton of very small
momentum in the lab system. 

For the reactions under consideration forward going charged pions, protons,
neutrons and deuterons have been detected in the forward detector and
identified by the $\Delta$E-E technique using corresponding information from
forward window, quirl, range and veto hodoscopes, 
respectively. Charged pions as well as gammas (from $\pi^0$ decay) have been
detected in the central part of WASA, which contains a solenoid magnetic field
with a mini drift chamber as well as the electromagnetic calorimeter. The
direction of neutrons could be measured in most cases by their hit pattern
resulting form recoil protons in forward and cental detectors.

With the exception of spectator protons and of neutrons the four-momenta of
all emitted particles have been measured. That way in all cases the full
events could be reconstructed with 1 -- 5 overconstraints and 
subjected to kinematic fits.

\section{Results}

\subsection{pp$\rightarrow NN\pi\pi$}

The two-pion production in pp-collisions has been measured from threshold up
to $T_p$=1.36 GeV. Total as well as differential cross sections have been
obtained for $pp\pi^+\pi^-$ \cite{JJ,WB,JP,iso}, $pp\pi^0\pi^0$ 
\cite{JJ,iso,TS}, $pn\pi^+\pi^0$ \cite{JJ,iso} and also $nn\pi^+\pi^+$
\cite{iso} channels. 

At low incident energies, i.e. $T_p\le$ 0.9 GeV, the data on the $\pi^+\pi^-$
and $\pi^0\pi^0$ channels have been successfully explained by excitation and
decay of the Roper resonance \cite{WB,JP,iso,TS}. However, the analysis of the
near-threshold $\pi^+\pi^-$ and $\pi^0\pi^0$ production, where the
differential observables are exceptionally sensitive to the interference between
the two Roper decay branches, provides  a ratio of approximately 1:1 for the
decay branching into $N\sigma$ and $\Delta\pi$ channels at a Breit-Wigner mass
of 1440 MeV \cite{TS,AE}. This branching ratio is in very good agreement with
the value obtained by the Bonn-Gatchina group in their partial-wave analysis 
\cite{boga}, but a factor of 4 larger than quoted in PDG \cite{pdg}, where a
ratio of 1:4(2)
 is quoted. Note
that at the pole mass for the Roper resonance, {\it i.e.} at m $\approx$ 1370
MeV, this branching ratio transforms even into 4:1, which points to a dominantly
monopole nature of the Roper excitation. 

At incident energies above 1 GeV, where the t-channel $\Delta\Delta$
excitation should take over, the data for the $\pi^+\pi^-$ and $\pi^0\pi^0$
channels change drastically. Indeed the $\Delta\Delta$ mechanism is identified
by observing the simultaneous excitation of $\Delta^{++}$ and $\Delta^0$ in
the appropriate $M_{p\pi^+}$ and $M_{p\pi^-}$ spectra. However, at the same
time we observe \cite{slov} a phase-space like behavior in the measured
$M_{\pi^+\pi^-}$ and $M_{\pi^0\pi^0}$ spectra rather than the predicted
\cite{luis} double-hump structure. Moreover, the predicted $\pi^0\pi^0$ total
cross section rises smoothly with increasing the incident energy, whereas the
measured total cross section levels off in the energy range 1.0-1.2 GeV before
it starts rising again for $T_p$ > 1.2 GeV \cite{iso}. This partial failure of
the theoretical description in the $\Delta\Delta$ region is presently still
under investigation. Possibly the $\rho$-exchange used in Ref. \cite{luis} is
the major reason for this failure. This is supported by the fact that the
$t$-channel $\Delta\Delta$ calculations in the framework of Risser and Shuster
\cite{ris}, which are based on pion-exchange solely, provide a superior
description of the data.
As an example Fig. 1 shows the invariant
mass spectra $M_{\pi^0\pi^0}$ and $M_{p\pi^0}$ from the  $pp \to pp\pi^0\pi^0$
reaction at $T_p$ = 1.3 GeV. 

\begin{figure}
\centering
\includegraphics[width=0.4\columnwidth]{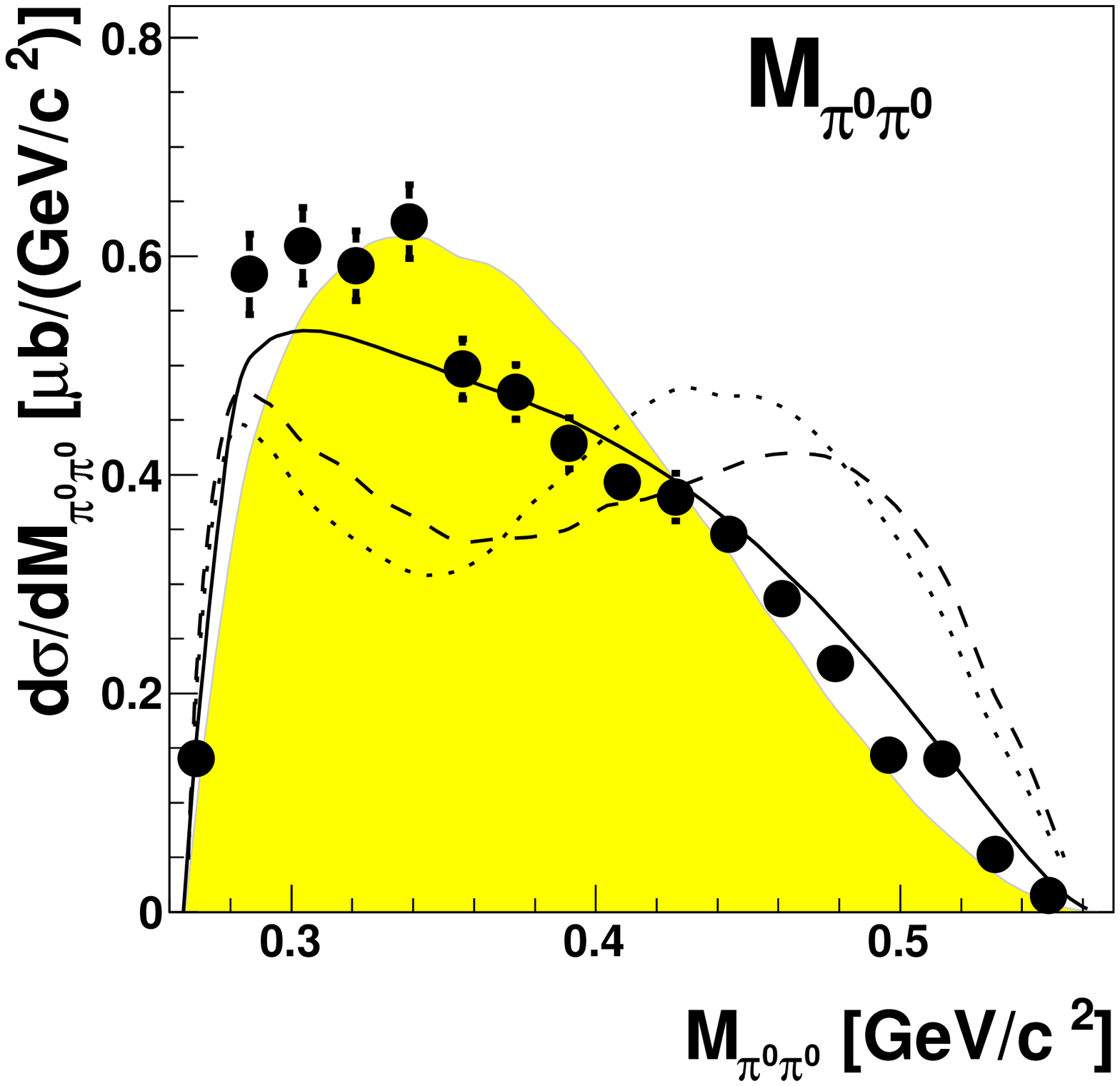}
\includegraphics[width=0.4\columnwidth]{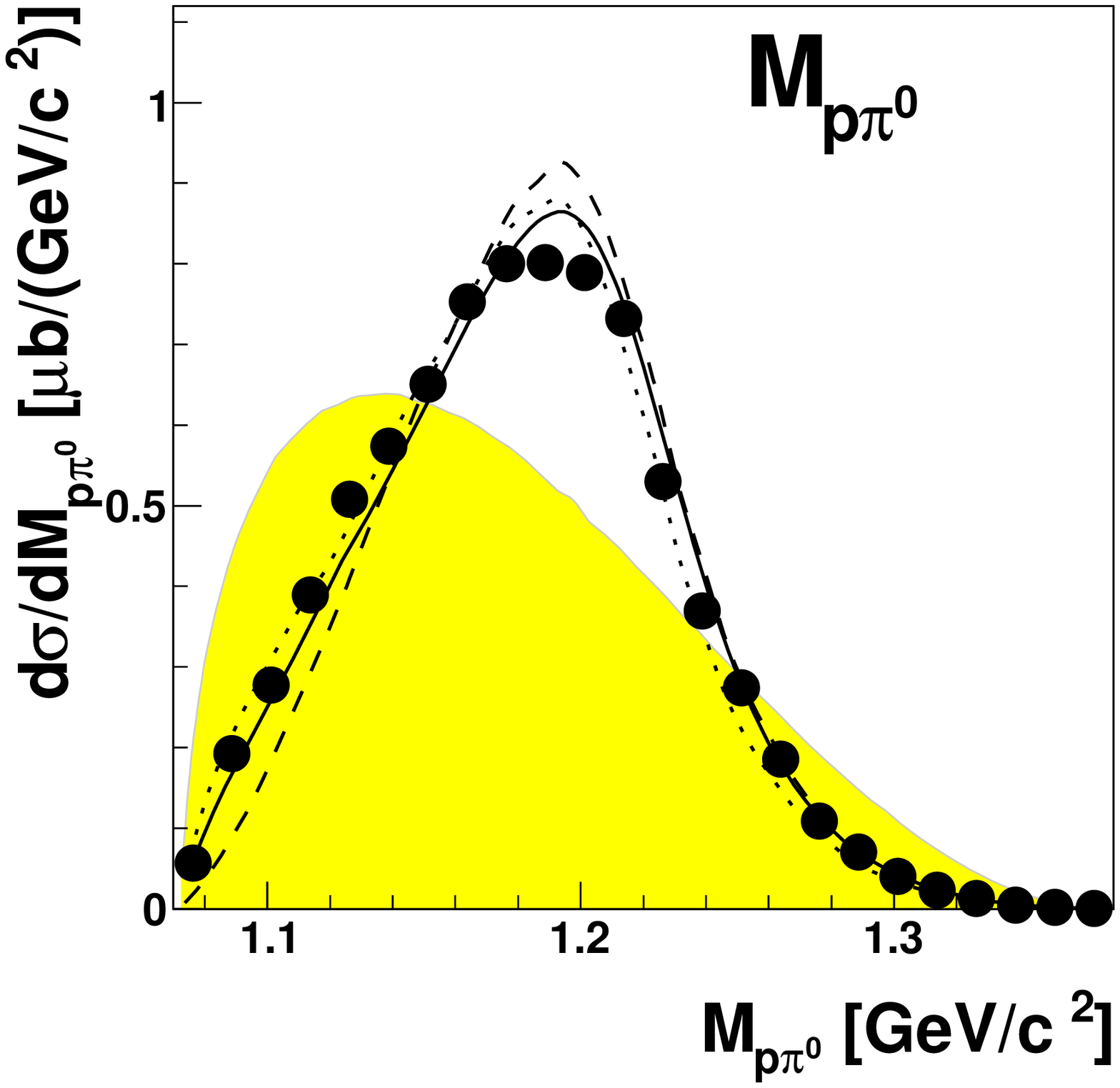}
\caption{Differential distributions of the invariant masses $M_{\pi^0\pi^0}$
   ({\bf left}) and $M_{p\pi}$ ({\bf right}) for the $pp \to pp\pi^0\pi^0$
  reaction at $T_p$ = 1.3 GeV. Solid dots represent the experimental
  results of this work. The shaded areas denote the phase space
  distributions. The dotted  curves show the original predictions of
  Ref. \cite{luis}, whereas the dashed lines give the same calculation,
  however, corrected for the proper Roper contributions as obtained from the
  analysis of differential data at lower beam energies \cite{WB,JP,TS,AE} and
  from isospin decomposition \cite{iso}. The solid lines give the $t$-channel
  $\Delta\Delta$ calculations in the framework of Risser and Shuster
  \cite{ris}, which is based on pion-exchange solely. All theoretical curves
  are normalized in area to the data.
}
\label{fig1}
\end{figure}

Another even more severe shortcoming of the calculations of Ref. \cite{luis}
is observed in the $nn\pi^+\pi^+$ channel, where the cross section turns out
to be a factor of 5 larger than predicted. Under the assumption that the 
$\Delta\Delta$ excitation is the dominant process in this channel, too, we would
expect the $\pi^0\pi^0$ cross section to be four times larger than the
$\pi^+\pi^+$ cross section by use of isospin relations. At $T_p$ = 1.1 GeV
we observe this ratio, however, to be close to unity \cite{iso}. We note that 
our results for the $\pi^0\pi^0$ and $\pi^+\pi^+$ cross sections are in
good agreement with previous bubble chamber data \cite{shim,eis}.

In order to be independent of model assumptions we have carried out an isospin
decomposition of the total cross sections \cite{iso}. As a result we obtained an
amplitude for the isoscalar $\pi\pi$ channel, which is in good agreement with
the theoretical predictions for the Roper excitation at energies $T_p <$ 1
GeV. For higher energies the theoretical cross section for the Roper excitation
process appears to be largely overestimated. 

Since the isotensor $\pi\pi$ channel cannot be described by just the
$\Delta\Delta$ excitation and small non- and semi-resonant contributions
\cite{luis}, contributions from higher-lying I=3/2 resonances have to be
considered as a possible way out \cite{iso}. The next higher-lying candidate
resonance is the $\Delta(1600)$. Since it has a large width of about 350
MeV and preferably decays via the $\Delta$(1232), it may contribute strongly to
the isotensor cross section and also to the isovector part. With the aid of
this excitation we obtain a good description of all 
$pp\rightarrow NN\pi\pi$ total cross section data up $T_p$=1.5 GeV.

\subsection{$NN\rightarrow ^2$He$\pi\pi$ and $NN \to d\pi\pi$}

If we constrain the relative momenta of the protons emitted in the $pp \to
pp\pi^0\pi^0$ reaction by a cut $M_{pp} < 2m_p +$ 10 MeV, then we select the
emitted 
$pp$ system to be at very small energy and in relative s-wave, {\it i.e.} in a
quasibound $^2$He final state. That way we obtain the transition to
double-pionic fusion. Indeed, if we apply this cut to our data, then we obtain
a double-hump structure in the $M_{\pi^0\pi^0}$ spectrum as predicted by the
$t$-channel $\Delta\Delta$ calculations of the early theoretical
interpretation of the ABC effect \cite{ris}. Also the recent COSY-ANKE data
\cite{anke} for the $pp \to ppX$ reaction with $M_{pp} < 2m_p +$ 3 MeV in a
restricted angular range are well accounted for by this process.

Having understood the transition from the two-pion production process for
unbound $NN$ systems to the quasibound one, we next investigate this process
to the 
really bound $NN$ system by studying the $pp \to d \pi^+\pi^0$ reaction. Since 
we have here the production of an isovector pion pair, Bose symmetry requires
this system to be in relative p-wave. This means, however, that the intensity at
small $M_{\pi^+\pi^0}$ must be suppressed and no low-mass enhancement is to be
expected. Indeed, our measurements of this reaction at $T_p$ = 1.1 GeV exhibit
no ABC effect and all differential data are in good agreement with the
conventional $t$-channel $\Delta\Delta$ process \cite{FK}. This process gives
also full account for the energy dependence of the total cross section of
this reaction, which exhibits a broad resonance-like structure of width
$\approx$ 230 MeV, {\it i.e.} twice the $\Delta$ width, and peaking at twice
the $\Delta$ mass. 

The situation gets drastically different, when switching from the isovector
$pp$ incident channel to the isoscalar $pn$ incident channel
\cite{hcl,MB,panic}. The total cross section of the $pn \to d \pi^0\pi^0$
reaction exhibits a strong and narrow resonance structure. In the very
recent high-statistics measurements with the WASA detector at COSY
\cite{panic} a width as narrow as 50 MeV is observed, 
{\it i.e.} nearly five times smaller than expected from the conventional
$t$-channel $\Delta\Delta$ process. Moreover this structure does not peak at
twice the $\Delta$ mass, but 80 MeV below it. In the region of this resonance
structure -- and only in this region -- the $M_{\pi^0\pi^0}$ spectrum exhibits
a  huge low-mass enhancement, much larger than expected from the conventional
$t$-channel $\Delta\Delta$ process. Also in contradiction to this process no
significant high-mass enhancement is found. Though these findings
are at variance with the conventional $t$-channel $\Delta\Delta$ process the
Dalitz plots clearly show the mutual excitation of two $\Delta$ states in the
course of the reaction.

\section{Summary and Conclusions}

For the two-pion production induced by $pp$ collisions data are now available
for all exit channels from threshold up to the region of the $\Delta\Delta$
excitation. Many of these data have been obtained by exclusive and
kinematically complete measurements, which provide both total and differential
cross sections. The bulk properties of these data are basically well
accounted for by $t$-channel processes leading to the associated production of
Roper and $\Delta(1600)$ resonances, respectively, or to  the mutual
excitation of both nucleons into the $\Delta$ state each. In the
$\Delta\Delta$ region we find some discrepancy between the observed
differential distributions and the calculations of ref. \cite{luis}. The
calculations in the framework of Risser and Shuster \cite{ris}, which only
account for simple pion exchange, give a superior description. This points to
a problem connected with $\rho$ exchange and/or short-range correlations as
handled in the calculations of Ref. \cite{luis}. 

Whereas the two-pion production induced by isovector $NN$ collisions appears
to be reasonably well understood by conventional $t$-channel processes, the
situation is fundamentally different in the two-pion production induced by
isoscalar $pn$ collisions. Unfortunately there are no or no high-quality data
for the non-fusion channels --- for obvious reasons, since they are very hard to
attack experimentally. However, there are now high-quality and high-statistics
data available for the quasi-free $pn \to d \pi^0\pi^0$ reaction both from
CELSIUS \cite{hcl,MB} and in particular now also from COSY \cite{panic}. In
both cases we find a huge low-mass enhancement in the $M_{\pi^0\pi^0}$ spectrum
(ABC effect), which is correlated with a narrow resonance-like structure in
the total cross section. From isospin relations for the conventional $t$-channel
$\Delta\Delta$ process, which has been demonstrated to work in the $pp$
induced two-pion production, we find that the latter process cannot account
for these intriguing features in the isoscalar $pn$ channel even in a
qualitative way.  
 
From these observations we are led to conclude that the observed
resonance must be due to an unconventional process, which proceeds through the
mutual excitation of two nucleons into the $\Delta\Delta$ system, possibly in
a kind of doorway mechanism. Due to the
antisymmetry condition for this system its quantum numbers can only be $I(J^P) =
0(1^+)$ or $0(3^+)$, if we assume the $\Delta\Delta$ system to be in relative
s-wave 
--- which again is very plausible, since the resonance is far below the nominal
$\Delta\Delta$ threshold. From the angular distributions we infer some evidence 
for the larger spin value. The detailed analysis of this matter is, however,
still in progress.

The situation that the ABC effect -- well-known now since 50 years -- is
associated with a narrow resonance structure in the total cross section comes
as a big surprise. It was not predicted by any theoretical considerations. If
true that this process proceeds via the $s$-channel, then we see here the
first manifestation of a genuine resonance in the dibaryon system --- as
predicted and searched for also since nearly 50 years.
We note that resonances in the baryon-baryon system, in particular in the
$\Delta\Delta$ system, actually have been predicted by various quark model
calculations \cite{ping,barnes,kam,oka,kuk,mot}.  

Finally we note that meanwhile we also carried out exclusive and kinematically
complete measurements of the double-pionic fusion to $^3$He and $^4$He. Again
we observe huge low-mass enhancements in the $M_{\pi^0\pi^0}$ spectra
\cite{mb,SK,AP}. This means that this resonance is obviously robust enough to
survive even in the nuclear medium. 

 
\section{Acknowledgements}

We acknowledge valuable discussions with L. Alvarez-Ruso, V. Anisovich,
D. Bugg, L. Dakhno, C. Hanhart, M. Kaskulov, V. Kukulin, E. Oset, A. Sibirtsev,
I. Strakovsky, F. Wang, W. Weise, C. Wilkin and the WASA-at-COSY collaboration
on this issue. 
This work has been supported by BMBF
(06TU201, 06TU261, 06TU9193), COSY-FFE (Forschungszentrum J\"ulich) and 
DFG (Europ. Graduiertenkolleg 683)



\end{document}